\documentstyle[prl,aps, epsfig]{revtex}
\begin{document}
\twocolumn[\hsize\textwidth\columnwidth\hsize\csname@twocolumnfalse\endcsname

\title{
Direct Imaging of the First Order Spin Flop Transition in the Layered 
Manganite $La_{1.4}Sr_{1.6}Mn_2O_7$} 
\author{
U. Welp, A. Berger, D. J. Miller, V. K. Vlasko-Vlasov, K. E. Gray, J. F. Mitchell} 
\address{
Materials Science Division, Argonne National Laboratory, Argonne, IL 
60439} 

\date{\today}
\maketitle
\begin{abstract}
Magnetic field induced transitions in the antiferromagnetic  
layered manganite $La_{1.4}Sr_{1.6}Mn_2O_7$ were studied using magnetization measurements 
and a high-resolution magneto-optical imaging technique.  We report the first direct 
observation of the formation of ferromagnetic domains appearing at the first order 
spin-flop transition. The magnetization process proceeds through nucleation of 
polarized domains at crystal defect sites and not by domain wall motion.  A 
small magnetic hysteresis is caused by the nucleation and annihilation of domains 
in the mixed state. These results establish a direct link between the magnetic 
structure on the atomic scale as seen in neutron scattering and the macroscopic 
properties of the sample as seen in magnetization and conductivity measurements.
\end{abstract} 
\narrowtext
\vskip1pc]

The naturally layered manganites of composition $La_{2-2x}Sr_{1+2x}Mn_2O_7$ have attracted 
much recent interest since in addition to the phenomenon of colossal magneto-resistance 
(CMR) \cite{1} they exhibit a variety of coupled magnetic, electronic and structural 
groundstates.  Depending on the doping level, x, the interplay of superexchange and 
double exchange interactions between the Mn-moments gives rise to competing 
antiferromagnetic and ferromagnetic spin arrangements \cite{2}.  For intermediate doping 
levels $0.5 > x > 0.32$, $La_{2-2x}Sr_{1+2x}Mn_2O_7$ is a planar ferromagnet with Curie temperatures, 
$T_c$, of about 100 K.  Concurrent with the magnetic order the $MnO_2$-bilayers become 
metallic being separated from each other by insulating $(La,Sr)_2O_2$ block layers.  
The coupling of the magnetic and electronic transitions has been attributed to the 
double-exchange mechanism \cite{1,3}.  Recent magnetization \cite{4} and neutron scattering \cite{5} 
experiments on the $x = 0.30$ - compound, $La_{1.4}Sr_{1.6}Mn_2O_7$, have shown that for this 
doping level the material is a type A antiferromagnet in which the ferromagnetic 
$MnO_2$ bi-layers are stacked antiferromagnetically along the $c$-axis with the Mn-moments 
oriented along the $c$-axis.  The unique feature of this material is the occurrence of 
a very large, anisotropic magnetoresistance at temperatures well below the magnetic 
ordering temperature which has been related to a field driven transition in the 
antiferromagnetic structure \cite{3,6}.

Here we present a study of the magnetic field induced transition using magnetization 
measurements and magneto-optical imaging.  Magneto-optical imaging allows for the first 
direct observation of the formation of magnetic domains and provides details of the 
local magnetization process which are unattainable from bulk magnetization and/or 
conductivity measurements.  At low temperature a first order spin-flop (metamagnetic) 
transition between the antiferromagnetic and polarized states is observed at a field 
of about 1.1 kOe applied along the $c$-axis.  For intermediate field 
values, we find direct evidence for a mixed state of 
coexisting spin-flop and antiferromagnetic domains as predicted in Ref. \cite{6}.  
The magnetization process in the mixed state proceeds through nucleation of spin-flop 
domains at crystal defect sites.  Domain wall motion appears to be of minor significance, 
especially in the onset region of the transition.  A small magnetic hysteresis, 
consistent with the first order nature of the transition, is caused by 
the difference between the domain nucleation and annihilation processes.

Single crystals of $La_{1.4}Sr_{1.6}Mn_2O_7$ were melt-grown in flowing 
20 \% $O_2$ (balance Ar) in a floating-zone optical image furnace.  The sample for the measurements described 
here was cleaved from the resulting polycrystalline boules \cite{7}.  It is shaped like 
an irregular plate (see Fig. 2) with lateral dimension of roughly $600 \times 
800 \mu m^2$ and thickness (along $c$) of $230 \mu m$.  The magneto-optical images were obtained using a 
high-resolution magneto-optical imaging technique originally developed for the study 
of superconductors \cite{8}.  It utilizes the Faraday rotation in an yttrium-iron-garnet 
film placed on top of the sample to create a map of the magnetic induction component 
normal to the imaged sample surface.

The magnetic characterization of our sample is summarized in Fig. 1. 
Figure 1a shows the temperature dependence of the magnetization measured in a field 
of 500 Oe applied along the tetragonal $c$-axis and perpendicular to this axis.  The 
pronounced anisotropy appearing at temperatures below 70 K with the $c$-axis 
magnetization dropping almost to zero is the typical behavior of a uniaxial 
antiferromagnet with the magnetic moments aligned parallel to $c$.  The peak in the 
$c$-axis data defines a Neel temperature of $T_N$ = 72 K.  The enhanced in-plane 
magnetization in the paramagnetic state has been observed before \cite{9} and is caused 
by ferromagnetic intergrowth phases commonly found in this layered material.  For 
our measurements we chose a fairly high field of 500 Oe as to minimize this 
contribution.

The field dependence of the c-axis magnetization at 20 K is shown in Fig. 1b for 
increasing ($M_{in}$) and decreasing ($M_{de}$) field.  The magnetization stays essentially 
zero (region ÒIÓ) until a critical field of $H_{SF} = 1.1 kOe$ is reached at which a 
steep, initially linear rise of the magnetization occurs.  The magnetization 
saturates near $H_S^{||} = 4.8 kOe$ at a value of 72 $emu/g$.  This magnetization 
behavior is indicative of a first order spin-flop transition at $H_{SF}$.  In the 
field region of 1.1 kOe to 4.8 kOe (region ÒIIÓ) a mixed state of coexisting 
antiferromagnetic and polarized domains is expected due to demagnetization 
effects \cite{10}.  The width of this region is given by the saturation magnetization 
and by the demagnetization coefficient which is about N = 0.55 for our sample 
geometry.  The spin arrangements in the various phases are indicated schematically 
in Fig. 1b.  Recently, a quantitative description \cite{6} of the conductivity and 
magnetization in terms of a model including in addition to double exchange 
first and second order superexchange and anisotropy, respectively, has been 
presented.
Figure 1b also shows the magnetization hysteresis $\Delta M = M_{de} - M_{in}$.
The magnetization process is essentially reversible, however, 
clearly resolved peaks of $\Delta M$ are observed near $H_{SF}$ and at 
$H_S^{||}$.  As discussed in more detail below these are caused by the nucleation of polarized and 
antiferromagnetic domains, respectively.

Figure 2 shows magneto-optical images of the normal component of the magnetic 
induction, $B_z$, at the sample surface. Bright contrast in these images corresponds 
to high local values of $B_z$.  Frames (a) to (e) correspond to the magnetization 
curve in Fig. 1b.  At applied fields below $H_{SF}$ (Fig. 2a) the sample is in the 
antiferromagnetic state and correspondingly there is no magnetic contrast in the 
bulk of the crystal.  Magnetic contrast exists only along the edges of the sample 
due to demagnetization effects, along a needle shaped domain extending to the 
lower right corner, and near the lower left corner.  The contrast near the lower 
left is associated with a non-magnetic inclusion labeled ÒaÓ in Fig. 2e.  
Electron microscopy identifies this inclusion as $(La,Sr)_2MnO_4$, a common second 
phase in the $La_{2-2x}Sr_{1+2x}Mn_2O_7$ manganites \cite{7}.  With increasing field, polarized 
domains seen as bright spots, nucleate.  These images show directly the coexistence 
of polarized and antiferromagnetic domains as expected for the mixed state of 
the first order spin-flop transition.

The magnetization process proceeds through the nucleation of new domains.  The expansion 
of polarized domains through continuous domain wall motion as seen in 
perpendicular magnetized ferromagnetic layers is not observed in this 
field range.  This behavior may have the following reasons. The thickness, 
$\delta$, of domain walls in traditional ferromagnets can be estimated from 
$\delta$ $\sim$ $\pi a (J_1/2K)^{1/2}$ where $J_1$ is the exchange energy, $K$ is the 
anisotropy energy and $a$ the lattice constant.  Using the available literature data 
for the in-plane exchange energy, $J_1$ $\sim$ $3.6\times 10^8 erg/cm^3$ \cite{11}, 
and $K$ $\sim$ $2\times 10^6 erg/cm^3$ \cite{6}, 
we estimate a domain wall thickness of about 30 lattice constants.  Such a small 
value implies that these domain walls can be pinned very effectively by 
imperfections in the underlying crystal structure which prohibits the occurrence 
of domain wall motion.

The smallest domains that are resolved have a diameter of about 5 $\mu m$.  
At sufficiently high density domains coalesce to form extended domain structures 
(see Figs. 2d, e) that exhibit clear striation along two orthogonal directions.  
Similar behavior has been observed on several crystals.  These striations coincide 
with the crystallographic (100) and (010) directions.  Across the dark wedge-shaped 
area highlighted by the two white lines in Fig. 2e) the direction of striation 
changes slightly.  This area coincides with a small angle grain boundary in the 
crystal structure.  These results indicate that the nucleation sites of the 
polarized domains are given by the underlying crystal defect structure.  Further 
evidence for this conclusion arises from the observation that the pattern of 
domains is exactly the same on successive field sweeps and that this pattern 
is independent of an applied in-plane field.  The nature of the nucleation sites 
and the cause for their patterning have not yet been determined.

Figure 3 shows field profiles along a horizontal line through the bright spot 
to the left of label Ò2Ó in Fig. 2e) for several applied fields.  With increasing 
field the profiles become increasingly complex.  In additon we observe that the 
local field can drop below the applied value (indicated by the dashed-dotted 
line for an applied field of 1680 Oe) in the vicinity of polarized regions.  
Such a field distribution which causes the formation of stripe domains in 
traditional magnets suppresses the nucleation of additional domains in the 
vicinity of existing polarized domains and thus stabilizes the striations.

Figures 2f to 2h show magneto-optical images after the sample surface has 
been polished down by about 20 $\mu m$.  The general domain pattern is not changed 
(compare Figs. 2d and 2g indicating that the dominating features are not a 
property of a specific surface structure but a property of the bulk.  
However, there are changes in the arrangement of small domains upon polishing.  
Figs. 2f and 2h show the domain pattern at 1176 Oe before and after the 
field has been ramped up to 1344 Oe (Fig. 2g).  The observed structures are 
almost identical, but there is a small hysteresis between the nucleation 
and annihilation of polarized domains, particularly visible near the top 
right corner.  This is the expected behavior at a first order transition 
and accounts for the positive magnetization hysteresis near $H_{SF}$ shown in 
Fig. 1c.  Similarly, the magnetization hysteresis near $H_S^{||}$ is caused by 
the hysteresis between nucleation and annihilation of antiferromagnetic 
domains in a predominantly polarized material.

Figure 4 shows local magnetization curves displaying the field dependence 
of $B_z$ in selected positions on the sample surface.  Here, position 1 and 2 
mark the bright spots to the left of labels Ò1Ó and Ò2Ó in Fig. 2e, and 
position 3 marks the dark spot to left of Ò3Ó in Fig. 2e, respectively.  
For Ò1Ó and Ò2Ó discontinuous jumps of $B_z$ by about 1000 G are observed 
consistent with the first order nature of the transition.  Up to an 
external field of 1680 Oe, the highest field reached for imaging, no 
spin-flop occurs in position 3.  Instead, the local field falls below 
the applied field due to the stray field of the surrounding polarized 
regions as described above.  Also included in Fig. 4 is the field dependence 
of sample-averaged $B_z$. In an applied field of 1680 G the field $(B_z - H)$ 
caused by the magnetization is about 230 G.  By solving the magneto-static 
surface integrals for a rectangular box with dimensions corresponding 
to our sample this field value can be converted into a magnetization of 
12 $emu/g$ which is in good agreement with the magnetometer measurement of 
14 $emu/g$ (Fig. 1b).  Similarly, the jump height of 1000 G can be converted 
into a saturation magnetization of 53 $emu/g$ which is about 25 \% smaller 
than expected.  This reduced value might be caused by a small gap between 
the sample surface and the magneto-optical film, typically several 
$\mu m$.  This will lead to a reduction of the measured value for $B_z$ above a 
domain by an error that increases as the lateral size of the domain 
decreases.  In addition a small, but finite closure angle between the 
Mn-moments in the spin-flop phase \cite{6} would cause a further increase of 
M in high fields that is not accounted for in the jump height.

The results shown in Fig. 2 can be directly related to measurements of 
the anisotropic magneto-conductivity of $La_{1.4}Sr_{1.6}Mn_2O_7$ \cite{6}.  The double 
exchange mechanism causes the intimate coupling of the electronic and 
magnetic state of this material.  In particular, a high c-axis conductivity 
is correlated with a ferromagnetic spin-alignment along the $c$-axis.  
Thus, the images shown in Fig. 2 can be seen as maps of the $c$-axis conductivity.  
Correspondingly, the sharp local steps of the magnetization (Fig. 4) imply 
extremely sharp steps in the conductivity which offers the potential for 
a field controlled resistive switch.  The in-plane transport occurs along 
a parallel circuit of ferromagnetic, highly conducting $MnO_2$ bi-layers in 
the antiferromagnetic and the fully polarized state.  At the spin-flop 
transition, however, domain walls start appearing and cause a local reduction 
in $\sigma_{ab}$.  As suggested in Ref. \cite{6} this mechanism gives a direct account 
of the minimum of $\sigma_{ab}$ observed near $H_{SF}$.

In conclusion, magnetic transitions in the antiferromagnetic state of 
$La_{1.4}Sr_{1.6}Mn_2O_7$  were studied using a high-resolution magneto-optical 
imaging technique.  We report the first direct observation of the formation 
of ferromagnetic domains appearing at the first order spin-flop transition. 
The magnetization process proceeds through nucleation of polarized domains 
at crystal defect sites, domain wall motion appears to be of minor 
significance, especially in the onset region of the transition.  A small 
magnetic hysteresis is caused by the nucleation and annihilation of domains 
in the mixed state.  These results establish a direct link between the 
magnetic structure on the atomic scale as seen in neutron scattering and 
the macroscopic properties of the sample as seen in magnetization and 
conductivity measurements.

This work was supported by the US DOE, BES - Materials Sciences under 
contract \#W-31-109-ENG-38.  We thank Qing'An Li and R. Osborn for 
helpful discussions.

\begin{figure}
\epsfxsize=3.4in
\epsffile{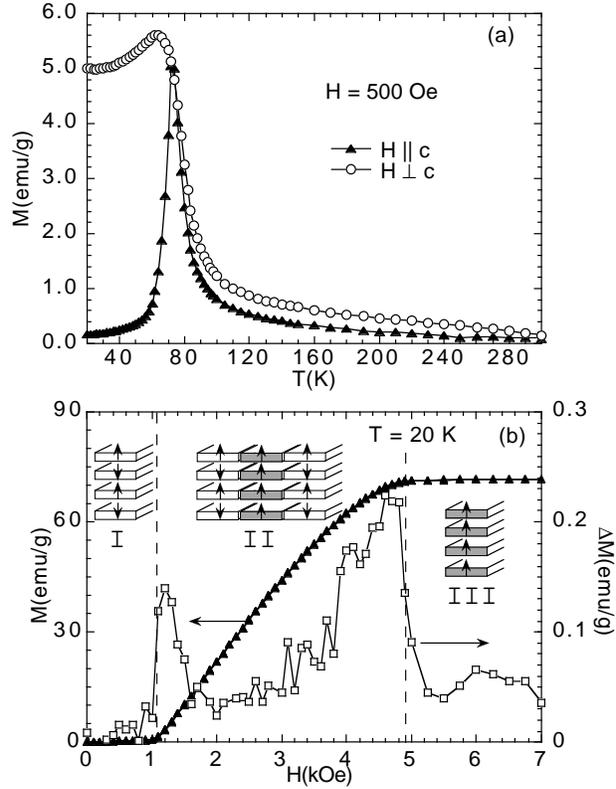}
\caption{ 
(a) Temperature dependence of the magnetization measured in a field 
of 500 Oe applied parllel and perpendicular to the $c$-axis.
(b) Field dependence of the $c$-axis magnetization at 20 K in 
increasing and decreasing field (left scale) and magnetization hysteresis 
$\Delta M$ (right scale) at 20 K.  Also shown are 
schematics of the spin arrangements in the antiferromagnetic ("I"), 
mixed ("II") and polarized ("III") phases.}
\label{fig1}
\end{figure}
 
\begin{figure}
\epsfxsize=3.0in
\epsffile{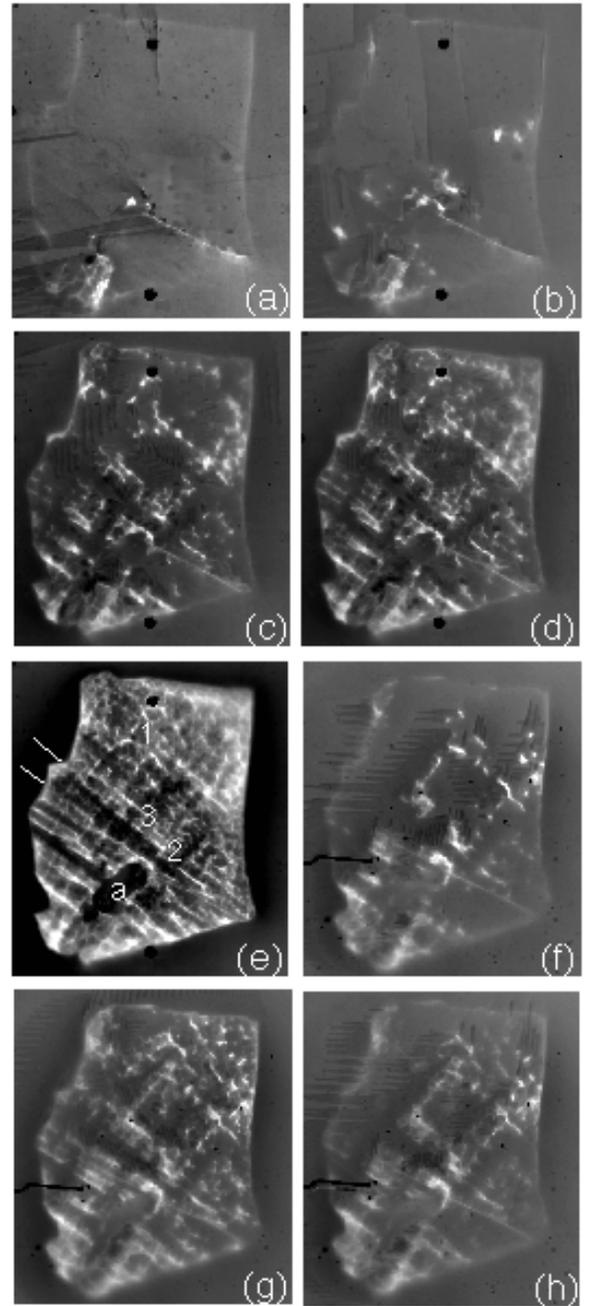}
\caption{ 
Magneto-optical images of the local magnetic induction, $B_{z}$, at 
the crystal surface taken at 20 K.  Images (a) to (e) are taken in 
1008 Oe (a), 1092 Oe (b), 1260 Oe (c), 1344 Oe (d), 1680 Oe (e).  In image 
(e) lable "a" indicates a $(La,Sr)_{2}MnO_{4}$ inclusion, "1", "2" 
and "3" indicate the positions of local magnetization curves (see 
Fig. 4).  The white lines mark a change in direction of the striations 
coinciding with a small angle grain boundary.  Images (f) to (h) are 
taken at 1176 Oe (f), 1344 Oe (g), 1176 Oe (h) after the sample surface 
has been polished by 20 $\mu m$.  A comparison (f) and (h) reveals the 
hysteresis between nucleation and annihilation of polarized domains.}
\label{fig2}
\end{figure} 

\begin{figure}
\epsfxsize=3.0in
\epsffile{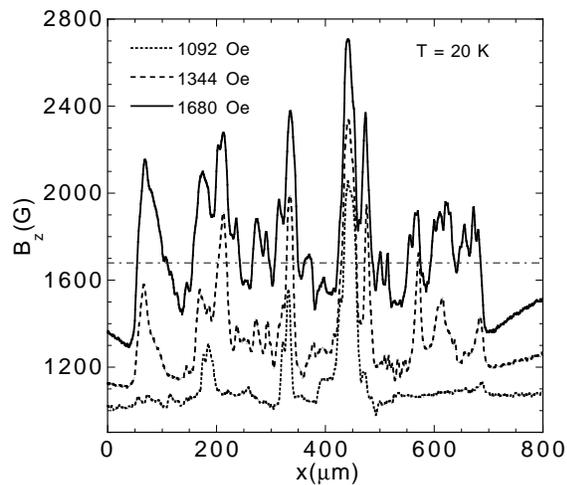}
\caption{ 
Field profiles along a horizontal line through position "2" in Fig. 
2e).  For 1680 Oe the applied field level is indicated by the dashed 
line.}
\label{fig3}
\end{figure} 

\begin{figure}
\epsfxsize=3.0in
\epsffile{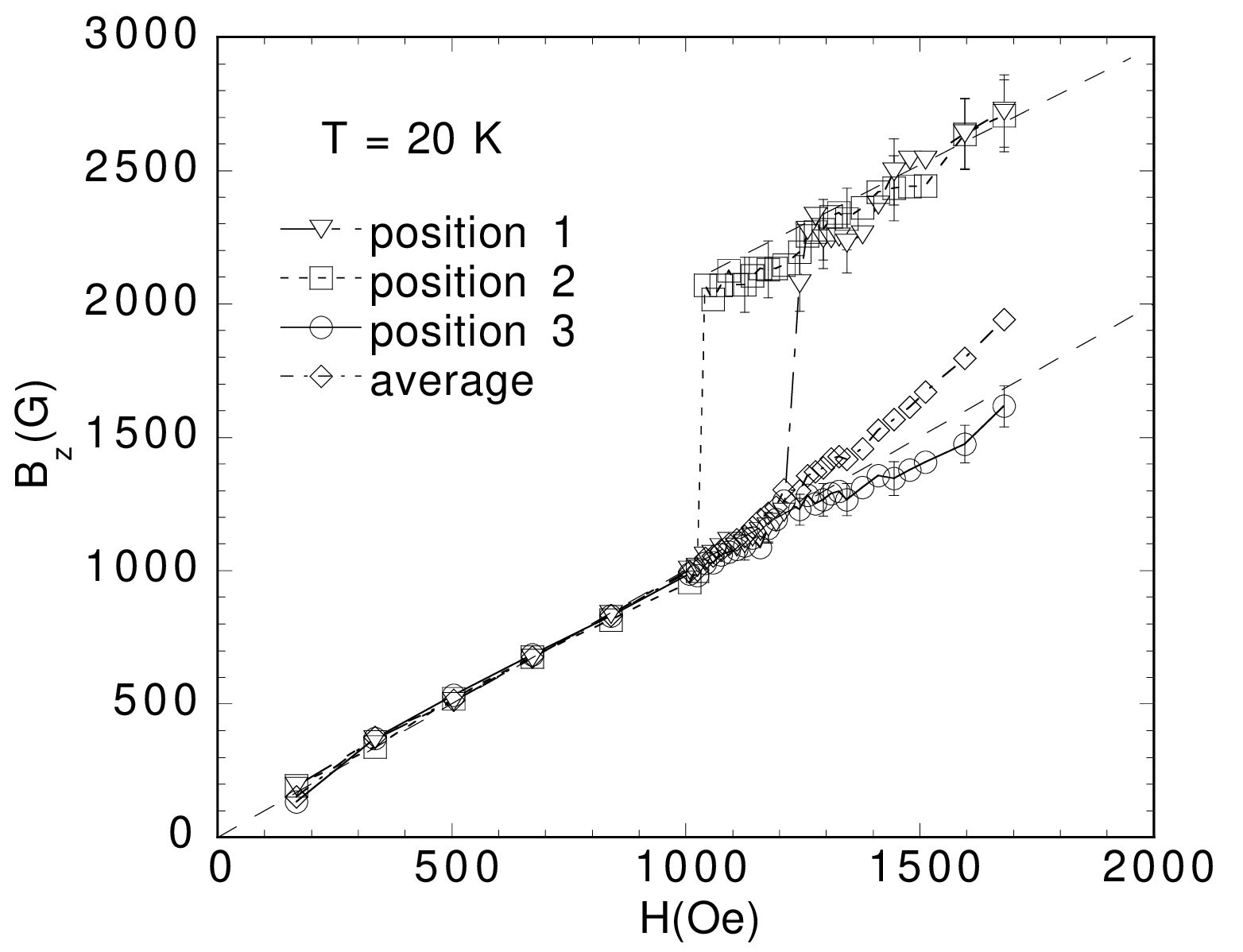}
\caption{ 
Local magnetization curves at the positions marked in Fig. 2e).  
Sharp, disontinuous jumps in the magnetization are observed at the 
transition into the spin-flop phase.  Also included is the average of 
$B_{z}$ over the sample surface.}
\label{fig4}
\end{figure}
 

\small 
\references
\bibitem{1}	T. Kimura et al., Science {\bf 274}, 1698 (1996). 
\bibitem{2}	M. Kubota et al., cond-mat/9902288.
\bibitem{3}	M. Imada et al., Rev. Mod. Phys. {\bf 70}, 1039 (1998);
J. B. Goodenough et al., in "Landolt - B\"ornstein" Vol. 4, part a, page 126.
\bibitem{4} T. Kimura et al., \prl {\bf 79}, 3720 (1997).
\bibitem{5}	T. G. Perring et al., \prb {\bf 58}, 14693 (1998); D. N. 
Argyriou et al., \prb {\bf 59}, 8695 (1999).
\bibitem{6}	Qing'An Li et al., cond-mat/9903452.
\bibitem{7}	J. F. Mitchell et al., \prb {\bf 55}, 63 (1997).
\bibitem{8}	L. A. Dorosinskii et al., Physica C{\bf 203}; for a recent 
review see: V. K. Vlasko-Vlasov et al., NATO ASI School "Physics and 
Materials Science of Vortex States, Flux Pinning and Dynamics", Kusadasi, 
Turkey, July 26 - August 8, 1998.
\bibitem{9}	S. D. Bader at al., J. Appl. Physics {\bf 83}, 6385 (1998); C. 
D. Potter et al., \prb {\bf 57}, 72 (1998).
\bibitem{10} B. E. Keen et al., J. Appl. Physics {\bf 37}, 1120 (1966); I. 
S. Jacob and P. E. Lawrence, Phys. Rev. {\bf 164}, 866 (1967).
\bibitem{11} S. Rosenkranz et al., to be published; H. Fujioka et al., 
cond-mat/9902253.  These experiments were performed on the x=0.4 - 
material which we consider a good estimate for the x=0.3 - compound.
\end{document}